\documentclass[12pt]{article}

\usepackage[numbers,square,sort&compress]{natbib}

\usepackage{comment}
\usepackage{adjustbox}
\usepackage{amscd,amsfonts,amsmath,amssymb,amsthm,bbm,bm,latexsym,mathrsfs}

\usepackage{enumerate}
\usepackage[shortlabels]{enumitem}

\usepackage[usenames]{color}
\usepackage{epsfig,epstopdf,graphics,graphicx,subfigure,longtable}
\graphicspath{{./figures/}}
\usepackage{tikz}
\usepackage{tkz-berge}

\usetikzlibrary{shapes,arrows,bayesnet}	% flow charts

\usepackage[toc,page]{appendix}
\usepackage[normalem]{ulem}% to highlight cancelled text
\usepackage{cancel}

\makeatother

\theoremstyle{remark}

\theoremstyle{plain}
\usepackage{nomencl}

\makenomenclature

\DeclareMathOperator{\diag}{Diag}

\begin{document}

%\bibliographystyle{natbib}

%%%%%%%%%%%%%%%%%%%%%%%%%%%%%%%%%%%%%%%%%%%%%%%%%%%%%%%%%%%%%%%%%%%%%%%%%%%%%%

  \title{\bf \large Comment on “Sparse Bayesian Factor Analysis when the Number of Factors is Unknown” by S. Frühwirth-Schnatter, D. Hosszejni, and H. Freitas Lopes written by Roberto Casarin and Antonio Peruzzi {\normalsize (Ca' Foscari University of Venice)}}
  \author{}
    
  \maketitle

\section{Introduction}

The techniques suggested in \citet{fruhwirth2024sparse}, \emph{FS-H-FL} hereafter, concern sparsity and factor selection and have enormous potential beyond standard factor analysis applications. We show how these techniques can be applied to Latent Space (LS) models for network data. These models suffer from well-known identification issues of the latent factors due to likelihood invariance to factor translation, reflection, and rotation (see \citet{hoff2002latent}). A set of observables can be instrumental in identifying the latent factors via auxiliary equations (see \citet{liu2021social}). These, in turn, share many analogies with the equations used in factor modeling, and we argue that the factor loading restrictions may be beneficial for achieving identification.

\section{Latent Space models}
Denote with $W = \{w_{ij},i,j = 1,\ldots,n\}$ the adjacency matrix of a weighted network $\mathcal{G}$, where the weights are integer-valued, $w_{ij} \in \mathbb{N}$. We assume the following model:
\begin{equation*}
    w_{ij} \overset{ind}{\sim} \mathcal{P}oi( \theta_{ij}),\quad \theta_{ij} = g(\alpha -  ||\mathbf{f}_{i} - \mathbf{f}_{j}||^2),
\end{equation*}
where $\mathcal{P}oi(\theta)$ denotes the Poisson distribution with intensity $\theta$, $g(\cdot): \mathbb{R} \rightarrow \mathbb{R}^{+}$ is a link function, $\alpha$ is an intercept parameter, $\mathbf{f}_{i}$, $i=1,\ldots,n$ is a collection of $d$-dimensional latent factors and $||\cdot||$ denotes the Euclidean distance. To avoid translation issues, one can assume $\sum_{i=1}^n f_{ik} = 0$ for $k = 1, \ldots,d$.

\begin{figure}[t]
    \centering
    \includegraphics[trim={0.5cm 0.5cm 0.5cm 0.8cm},clip,width=0.68\linewidth]{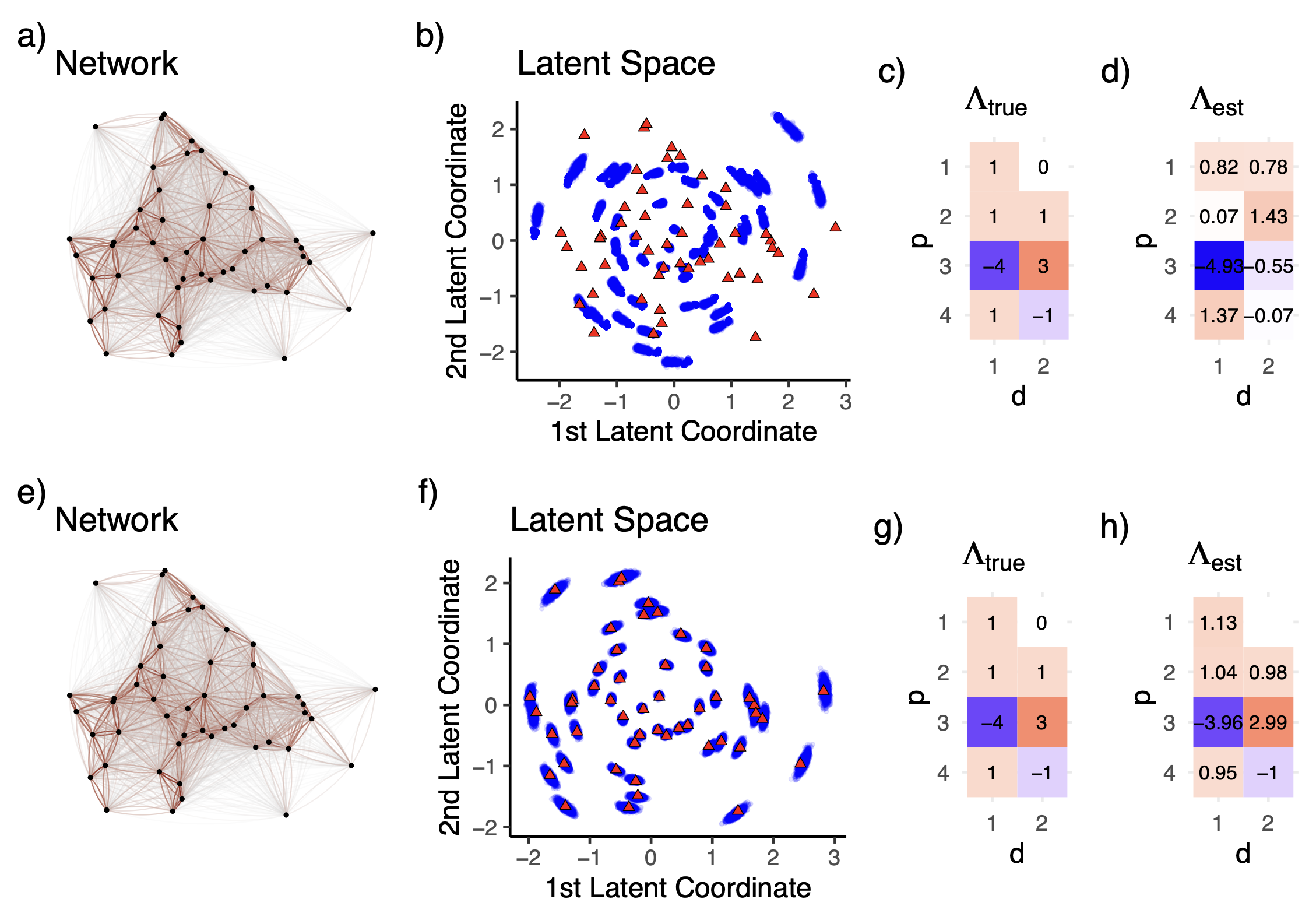}
    \caption{Results for an LS model without and with restrictions (top and bottom, respectively). Panel a) and e) 
    report the observed network width edge gradient proportional to the absolute distance between the observed and predicted weight (darker edge colors).  Panel b) and f) report the posterior draws (blue dots) against the true latent coordinates (red triangles). The true value of $\Lambda$ is in Panel c) and g). Panel d) and h) report the posterior means of $\Lambda$ without and with PLT restrictions, respectively.}
    \label{fig:enter-label}
\end{figure}

The latent factors can be interpreted via a set of node-specific observables $Y$ with the following interpretation factor model:
\begin{equation*}
        Y = \Lambda \mathbf{f}  + \boldsymbol{\varepsilon},\quad \boldsymbol{\varepsilon}\sim\mathcal{MN}_{p,n}(O, \Sigma_p, I_n),
\end{equation*}
where $Y$ is an $p \times n$ matrix of interpretation variables, $\mathbf{f} = (\mathbf{f}_1, \mathbf{f}_2, \ldots,  \mathbf{f}_n)$ is a $d \times n$ matrix obtained by stacking the factors, $\Lambda = (\boldsymbol{\lambda}_1, \boldsymbol{\lambda}_2, \ldots,  \boldsymbol{\lambda}_d)$ is a $p \times d$ matrix of loadings with $\boldsymbol{\lambda}_k = (\lambda_{1k}, \lambda_{2k}, \ldots, \lambda_{lk}, \ldots, \lambda_{pk})'$ and $\boldsymbol{\varepsilon}$ is a $p \times n$ matrix of independent normal error terms with $\Sigma_p = \diag(\sigma^2_{1},\ldots, \sigma^2_{p})$.
 We are interested in achieving \emph{row sparsity} for $\Lambda$. Similarly to \emph{FS-H-FL}, we assume the following prior distributions:
\begin{align*}
&\alpha \sim\mathcal{N}(0, \sigma^2_\alpha ),\quad
\mathbf{f}_{i} \sim \mathcal{N}_d\left(\boldsymbol{0},(1-1/d)^{-1}I_d\right), \quad\sigma^2_i \sim  \mathcal{IG}\left(c_0, C_0\right),
\\
&\tau_l \sim \mathcal{B}e\left(1, 1 \right),\quad \sigma_k^2 \sim \mathcal{IG}\left(c_\sigma, b_\sigma\right), \quad \kappa \sim \mathcal{IG}\left(c_\kappa, b_\kappa\right),\\
&\lambda_{lk} \mid \kappa, \sigma_k^2, \tau_l \sim\left(1-\tau_l\right) \delta_0+\tau_l \mathcal{N}\left(0, \kappa \sigma_k^2\right).\\
\end{align*}
Figure \ref{fig:enter-label} presents the posterior results for an LS model with $d = 2$ and $p = 4$ for the unrestricted and restricted $\Lambda$ (top and bottom panels, respectively). Panel b) shows the identification issue, and Panel f) the effectiveness of the restrictions on $\Lambda$ to achieve identification of the set of latent factors $\mathbf{f}$. The factor identification is obtained via PLT restriction, i.e. $\lambda_{kk} > 0$ and $\lambda_{lk} = 0$ for $k>l$.  As discussed in \emph{FS-H-FL}, the PLT structure may be too restrictive. Therefore, we speculate on imposing an ordered or unordered GLT structure on $\Lambda$.

\section{Conclusion}
As further research, we suggest extending the authors' approach to nonlinear factor models. This is a stimulating work, and we are therefore very pleased to be able to propose the vote of thanks to the authors for
their contribution.

%%%%%%%%%%%%%%%%%%%%%%%%%%%%%%%%%%%%%%%%%%%%%%
%% Acknowledgements                         %%
%% should be provided in the                %%
%% Acknowledgements section.                %%
%%%%%%%%%%%%%%%%%%%%%%%%%%%%%%%%%%%%%%%%%%%%%%

%%%%%%%%%%%%%%%%%%%%%%%%%%%%%%%%%%%%%%%%%%%%%%
%% Funding information, if any,             %%
%% should be provided in the                %%
%% funding section.                         %%
%%%%%%%%%%%%%%%%%%%%%%%%%%%%%%%%%%%%%%%%%%%%%%

\section*{Acknowledgements}
This discussion was supported by the EU -  NextGenerationEU, in the framework of the GRINS - Growing Resilient, INclusive and Sustainable project (GRINS PE00000018 - CUP H73C22000930001), National Recovery and Resilience Plan (NRRP) - PE9 - Mission 4, C2, Intervention 1.3. The views and opinions expressed are solely those of the authors and do not necessarily reflect those of the EU, nor can the EU be held responsible for them.

\bibliographystyle{agsm}
\bibliography{sample}

\end{document}